\documentclass[twocolumn,showpacs,preprintnumbers,amsmath,amssymb]{revtex4}

\usepackage{graphicx}
\usepackage{amsmath}
\DeclareMathOperator{\Tr}{Tr}
\DeclareMathOperator{\Sign}{Sign}
\begin{document}

\title{ \bf 
Diagrammatic Quantum Monte Carlo Algorithm in Momentum Representation:
Hess-Fairbank Effect and Mesoscopics in 1D BEC with
Attractive Interaction}

\author{P.~F.~Kartsev\/\thanks{kartsev@kurm.polyn.kiae.su}}
\affiliation{
Department of Solid-State Physics, 
Moscow Engineering-Physics Institute (State University),
Kashirskoe sh. 31, 115409 Moscow, Russia
}

\begin{abstract} 
A novel algorithm of Diagrammatic Quantum Monte Carlo
in momentum representation is reported in details.
New models can be studied with this algorithm.
For Bose systems with attractive interaction,
the algorithm is free of the well-known minus-sign problem,
while in other models it is weaker than in real-space methods.

Using this algorithm,
we present the results of an exact numeric simulation of $N$
one-dimensional bosons with attractive $\delta$-functional
interaction in a rotating ring. We prove that in the large-$N$
limit the system can be described by conventional methods of
weakly interacting gas, the dimensionless parameter of weak
interaction being just $1/N$. When the strength of interaction is
less then a certain threshold value, the dependence of angular
momentum on the rotation frequency features plateaus
characteristic of the irrotational fluid (the Hess-Fairbank
effect).

\end{abstract}


\maketitle

\section{QMC algorithm in momentum representation against sign problem}
\label{algor}

\subsection{Background: sign problem}
The Quantum Monte Carlo (QMC) methods
(variational, determinant, trajectory \textit{etc}.)
 prove their usefulness
for studying thermodynamics of diverse quantum systems.
In QMC, 
the partition function $Z = \Tr{e^{- \beta \hat H}}$
is broken into series and summed
using importance sampling.
Each term is represented by unique set
of inner parameters (MC configuration)
and can be positive or negative depending on
the actual MC configuration,
the model Hamiltonian, and particle statistics.
Alternating sign results in fluctuation of
partition function and other quantities used in calculation.
As the temperature is lowered, the errors grow larger
making the calculation sluggish or even impossible.
This is the so-called minus-sign problem,
the inherent feature of most trajectory QMC methods.

Some cluster methods are free of minus sign.
The determinant QMC method \cite{det_qmc} is free of minus sign,
but unfortunately, its application is restricted by
fermionic and spin systems.
Moreover, its running time as a function of cluster size $L$
scales as $L^3$ while in trajectory  methods it is linear by $L$. 
So for large clusters, the expected slowing down 
of determinant method is stronger than that
of trajectory method caused by sign problem.
The method of exact diagonalization of Hamiltonian\cite{exact_diag}
is more useful and applies to wider class of models.
Nevertheless, the size of system
is limited by $L_{\mathrm{max}} = 10 \div 12$
as the amount of needed calculations grows exponentially with 
increasing $L$.

Some sort of correction\cite{e0_corr} makes possible to determine
the ground-state energy with good precision
even if approaching the temperature low enough
is prevented by sign problem.
For electrons on simple square (or cubic) lattice
and hopping only to nearest neighbours,
the particle-hole transormation helps to remove
part of sign not linked with Fermi statistics.
But generally, sign problem is present in simulation.

This Letter is organized as follows.
In Section \ref{algor},
we present the trajectory algorithm in momentum representation
developed in the framework of Diagrammatic QMC method
weakening or removing minus-sign in many models.
In Section \ref{applied}, we study the 
one-dimensional bosonic system with attractive interaction
in a rotating ring
taken by Ueda and Leggett as a model of irrotational fluid
\cite{ul}.
This model can not be simulated
with usual real-space methods,
in part due to sign problem,
while the algorithm described in this Letter
is very efficient in this system
allowing to simulate up to 100 and more particles.
\subsection{DQMC basics}

In this paragraph, we repeat shortly the DQMC basics
to be used later \cite{ctwl}.

\paragraph*{Decomposition.}
Diagrammatic QMC is based on the following
decomposition of partition function
into series of interaction representation:
\begin{eqnarray}
\label{Z_dqmc}
\nonumber
Z = 
\sum \limits_{m=0}^{\infty} 
\sum \limits_{\substack{\{n^{(1)} \} \\ ... \\ \{n^{(m)} \} }} 
\int \limits_0^\beta (-d\tau_{m-1})
\int \limits_0^{\tau_{m-1}}  (-d\tau_{m-2})
...
\int \limits_0^{\tau_1}(-d\tau_0) 
\times
\\
\times
e^{- \beta E_0^{(0)}}
\prod \limits_{j=1}^{m}
e^{- \tau_{j} E_0^{(j-1)}}
\left< \{ n^{(j-1)} \} \right| \hat V \left|
\{ n^{(j)} \}  \right>
e^{\tau_j E_0^{(j)}
}
,
\end{eqnarray}
where
 the system Hamiltonian is $\hat H = \hat H_0 + \hat V$,
energies $E_0(...)$ are given by $\hat H_0$
which is chosen to be diagonal on
occupation numbers $\{ n \}$,
 $\tau_m \equiv \tau_0 + \beta$, $\{n^{(m)}\} \equiv \{n^{(0)} \}$,

Each term in Eq. (\ref{Z_dqmc}) is represented by
its own picture of particle trajectories
in $({\bf x}, \tau)$-space, where $\tau = 0 \div \beta$,
$\beta = 1/k_B T$.
The weight of a given MC configuration writes as
\begin{equation}
\nonumber
W_{\mathrm{MC}} \sim \prod
\limits_{j=1}^{m}
 \left( -\Delta\tau 
\left< \dots \right|
 \hat V \left| \dots \right> \exp{(...)}\right).
\end{equation}
Then $\tau_i$ mark the points
of worldline distortions, the so-called ``kinks''
\cite{ctwl}.
The imaginary-time step $\Delta \tau$ 
can be taken small enough ($\sim 10^{-8} \beta$).

\paragraph*{Process of calculation.}
The importance sampling (Metropolis algorithm)
consists in random transformations of
MC configurations obeying the following requirements:
\begin{itemize}
\item{ \textit{Full set.}}
Two arbitrary MC configurations with nonzero weight
can be transformed into each other
with nonzero probability
and in finite number of steps;
\item{ \textit{Balance.}}
For each updating process transforming MC configurations
A to B  (``direct'' process),
we juxtapose the ``inverse'' process
transforming MC configurations B exactly to A.
Additionally, their frequencies must meet the balance equation
\begin{equation}
\nonumber
\label{eq_balance}
|W_A| P^{\mathstrut}_{\rightarrow} p^{(\mathrm{acc}) \mathstrut}_{\rightarrow}=
 |W_B| P^{\mathstrut}_{\leftarrow} p^{(\mathrm{acc}) \mathstrut}_{\leftarrow},
\end{equation}
with $W_A$ and $W_B$ weights of MC configurations A and B,
$P_{\rightarrow}$ and $P_{\leftarrow}$
the frequencies to call direct and inverse processes,
and
$p^{(\mathrm{acc})}_{\rightarrow}$ and
$p^{(\mathrm{acc})}_{\leftarrow}$
the probabilities to accept
the respective update.
The latter are usually determined
from the relations
\begin{align*}
 & p^{(\mathrm{acc})}_{\rightarrow} = \alpha, 
  && p^{(\mathrm{acc})}_{\leftarrow} = 1,
   &&& \text{if } \alpha \le 1, \\
 & p^{(\mathrm{acc})}_{\rightarrow} = 1,
  && p^{(\mathrm{rej})}_{\leftarrow} = 1 / \alpha, 
   &&& \text{if } \alpha > 1, \\
\end{align*}
where
$\alpha \equiv
\left |\frac{W_B P_{\rightarrow} }{ W_A P_{\leftarrow}} \right|$
is the so-called ``acceptance ratio''.
Therefore, to determine the probability of accepting the update,
we must know parameters of respective inverse process.

\end{itemize}

There remain some freedom of choosing the time $\tau$
of newly-created (or shifted) kink.
The most reasonable approach is to choose $\tau$
with probability
$\frac{\Delta \tau e^{-\delta E (\tau - \tau_{min})}}
{\mathrm{Z_1}(\delta E, \tau_{max}-\tau_{min})}$
using relation
\begin{equation}
\label{tau}
 \tau = \tau_{min} + \mathrm{P} (\Delta E, \tau_{max}-\tau_{min})
\end{equation}
(see Appendix \ref{Z1}),
 according to the fact that the weight of new MC configuration
is proportional to $e^{- \Delta E \tau}$.

\subsection{DQMC in momentum representation}

The algorithm described here is developed
for systems with a Hubbard-like Hamiltonian
in momentum representation
\begin{equation}
\label{ham}
\hat H = \sum \limits_{p} 
{ \epsilon^{\mathstrut}_p
\hat a^{\mathstrut +}_p
\hat a^{\mathstrut}_p
}
 + \sum \limits_{p,q,r,s} { U^{\mathstrut}_{pqrs}
 \hat a^{\mathstrut+}_{p} \hat a^{\mathstrut +}_{q}
 \hat a^{\mathstrut}_{r} \hat a^{\mathstrut}_{s}}.
\end{equation}

Though typically, interaction term conserves full momentum
$U_{pqrs} = U_{pq} \delta_{p+q=r+s}$,
but it is not mandatory for this algorithm.

\begin{figure}
\begin{center}
\includegraphics[width=8cm]{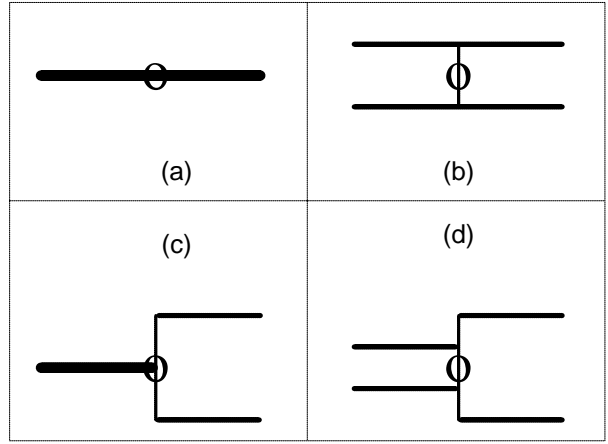}
\end{center}
\caption{
Kinks generated by two-particle interaction:
\textit{a, b} -- diagonal, \textit{c, d} -- non-diagonal;
\textit{a, b}
in case of contact interaction $U_{pqrs} = U_0 \delta_{p+q=r+s}$
 can be taken into account analytically;
\textit{a, c} appear in simulation of Bose systems only.
Here and troughout the paper,
imaginary-time axis is plotted horizontally,
momenta are placed vertically,
occupation number indicated by line thickness.
}
\label{kinks}
\end{figure}

\paragraph*{Kinks.}
The interaction term taken as a perturbation,
generates four-ended kinks shown in Fig.~\ref{kinks}.
The multiplier entering the configuration weight
due to each kink,
is given by sum of respective
 $U_{pqrs}\sqrt{n_p n_q n_r n_s}$ for
all nonidentical recombinations of momenta $p$,$q$,$r$,$s$.
For Fermi systems, half of these terms get negative sign. 
This allows to use the standard relation
$\Sign{(F)} = (-1)^{\sum \limits_i {\left( W_i^{(\tau)} -1 \right)}}$
linking fermionic sign of configuration
and time winding numbers $W_i^{(\tau)}$ of worldlines
which holds in real-space trajectory methods.

\begin{figure}
\begin{center}
\includegraphics[width=8cm]{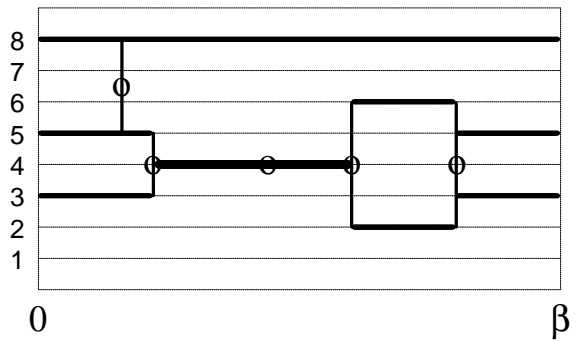}
\end{center}
\caption{
Sample Monte Carlo configuration with all types of kinks.
}
\label{examp}
\end{figure}

The example of MC configuration is sketched
in Fig.~\ref{examp}.

\paragraph*{Sign problem.}
The sign problem in usual real-space QMC algorithm
is caused by hopping term. In this algorithm,
on the contrary, it is caused by sign of interaction.
Thus, weakly-interacting systems can be simulated
in momentum representation
with much greater precision.

For attractive interaction, the algorithm is free
of sign problem, as every kink enters
MC weight with positive factor.
The particle-hole transformation for electrons on a lattice
can remove part of sign not linked with Fermi statistics,
in case when their interaction is
$\sum \limits_{ij} V_{ij} n_{i \uparrow} n_{j \downarrow}$.

In simple but rather wide case of point interaction
$U_{pqrs} = U_0 \delta_{p+q,r+s}$,
the diagonal part of interaction (kinks Fig.~\ref{kinks}~(a,b))
is summed analytically
and the MC configurations with single kink giving most part of sign problem,
 become impossible.
Monte Carlo weight can become negative for three and more kinks,
but in lower orders by $\beta U$ all diagrams are positive-definite
with no relation to particle statistics.
As a result, the system can be simulated at lower temperature.

\paragraph*{Updates.}

\begin{figure}
\begin{center}
\includegraphics[width=8cm]{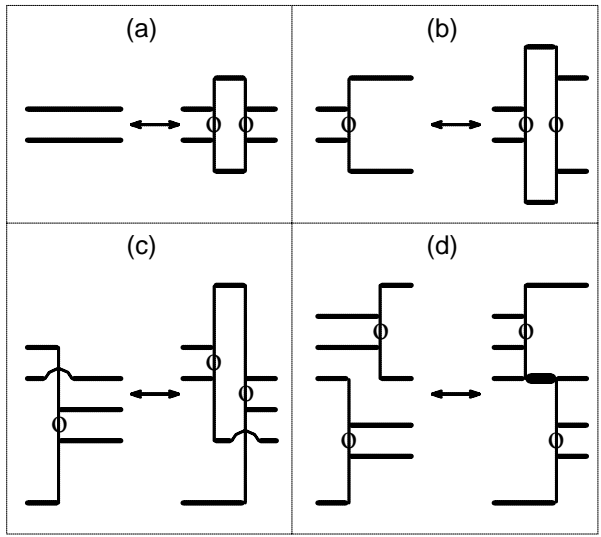}
\end{center}
\caption{
The updating processes for non-diagonal kinks:
Creation/Annihilation of Two Kinks (a),
Expanding/Contracting of a Kink (b),
Entangling of a Kink with a Worldline (c),
and Shifting of a Kink Through Another (d).
}
\label{procs}
\end{figure}

The processes chosen to update 
worldline configuration,
are shown in Fig.~\ref{procs}.
In addition, the time-shifting of kink should be used
to speed-up calculation.
For susbsystem of diagonal kinks, their creation/annihilation 
and time-shifting are enough.

\begin{figure}
\begin{center}
\includegraphics[width=8cm]{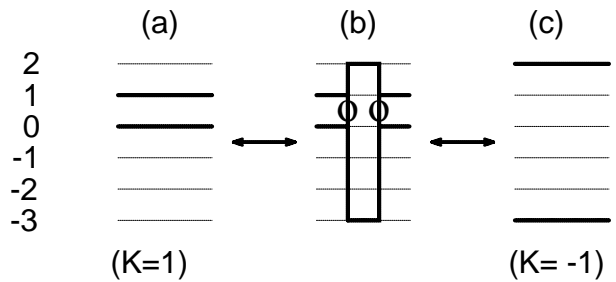}
\end{center}
\caption{
Creation of two ``biased'' kinks 
and annihilation of them in different way
change full momentum of the system.
}
\label{bias}
\end{figure}

Note, in case of momentum conservation $p+q = r+s$
these processes can not change full momentum $K$ of the system.
As a result, the calculation is done in the sector
of phase space with fixed $K$.
This situation can be corrected by introducing fictitious kinks 
changing number of particles ($\eta \sum \limits_i
(\hat a_i^{\mathstrut} + \hat a_i^{\mathstrut +})$,
like in Worm algorithm\cite{worm}),
 or full momentum
($\eta \sum \limits_{\substack{pqrs \\ p+q \ne r+s}} 
 \hat a^{\mathstrut+}_{p} \hat a^{\mathstrut +}_{q}
 \hat a^{\mathstrut}_{r} \hat a^{\mathstrut}_{s}$, 
see Fig.~\ref{bias}), with $\eta$ small enough.
However, fictitious kinks are not needed when calculating energy levels
characterized by their own value of full momentum.

\subsection{Applications of the method}

\paragraph*{1D Fermi Hubbard model.}
In testing purposes, we used the algorithm
 to calculate the ground state
of the one-dimensional fermionic system with the Hamiltonian
$
\hat H = t \sum \limits_{<ij> \sigma} {( \hat a^{\mathstrut +}_{i \sigma}
\hat a^{\mathstrut}_{j \sigma} + H.c.)}
+ U \sum \limits_i { \hat n_{i \uparrow} \hat n_{i \downarrow}}
$
with t=1, U=-1, length of the chain $L=8$,  $K=0$,
 number of particles $N_{\uparrow}=N_{\downarrow}=4$.
The result $E_0 \simeq -11.9523$
was checked with exact diagonalization ($E_0 = -11.952326$)
and real-space Worm algorithm.

\begin{figure}
\begin{center}
\includegraphics[width=8cm]{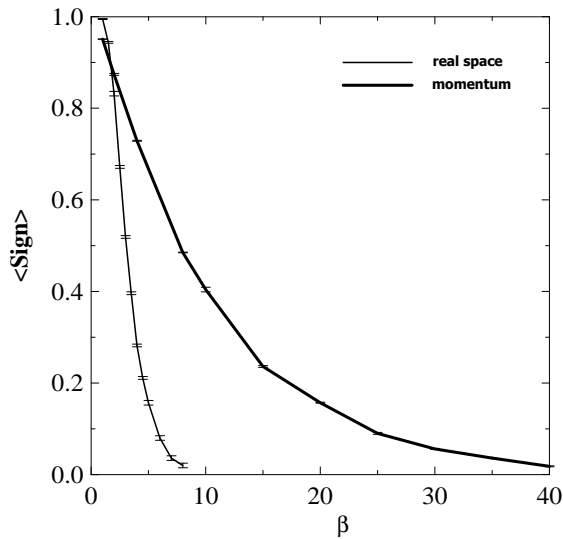}
\end{center}
\caption{Average sign of MC configuration 
as a function of $\beta = 1 / k_B T$ 
for QMC calculations of
the 1D Fermi Hubbard model
with $L = 8$ sites, $N_\uparrow = N_\downarrow=4$, $U=-1$, $t=1$,
in both real space and momentum representations.
The calculations become impracticable 
for $\langle Sign \rangle < 0.01$, 
so with momentum representation (and fixed full momentum $K=0$)
$\beta_\mathrm{max}$ increases from 7 to 40.
}
\label{sgn_1d}
\end{figure}

The average sign
for both QMC algorithms
as a function of $\beta$
is shown in Fig.~\ref{sgn_1d}.
The graph confirms our assumption
that sign problem is much weaker in new algorithm
even for Fermi systems.
With new algorithm, the maximal possible value of $\beta$ 
in sample system is increased from 7 to 40.

It is worthy to note that such rather good precision of 5 digits
became possible because of fixation of full momentum $K=0$.
The second energy level $E_1 = -11.901727$ 
corresponding to $K \ne 0$
is extremely close to the ground state,
so the precision of real-space QMC is limited by 3 digits
as lowering of the temperature is prevented by sign problem.

\paragraph*{Fractional Quantum Hall Effect.}
In studying Wigner Crystallization in 
Fractional Quantum Hall Effect,
the fermionic Hamiltonian of view (\ref{ham}) was obtained
and then analysed using exact diagonalization method \cite{fqhe}.
Now the QMC algorithm described here, was applied
to this system in order to confirm or deny
the existence of Wigner crystal.
Unfortunately, the fermionic sign problem appeared too strong
to determine the ground-state properties in this model.

\paragraph*{Bose gas with attraction in a box.}

The best model for the algorithm described in this Letter
is a Bose gas in a box with attractive interaction.
In addition to the freedom of sign problem,
this algorithm gets more advantage here,
as the model can not be simulated by real-space QMC
without errors \cite{bose1d_error}.
These are caused by discretization of
continuous real space needed to apply
lattice QMC methods.

\section{Rotating Bose-Einstein Condensate
with Attractive Interaction in One Dimension:
Hess-Fairbank Effect and Mesoscopics}
\label{applied}

\subsection{Background: macroscopic study}

Recent remarkable progress in Bose-Einstein condensation of dilute
alkali gases \cite{BEC} has opened up an opportunity of studying
delicate quantum phenomena in ultracold multi-atomic systems. One
of intriguing set-ups is a system of (quasi-)one-dimensional (1D)
bosons with attractive interaction -- like $^7$Li -- in a rotating
ring \cite{ul,smerzi}. In contrast to 3D case, where the
Bose-Einstein condensate of attracting atoms becomes unstable with
respect to a collapse \cite{bosenova} above a certain threshold
[which decreases (vanishes) with decreasing (vanishing) the
trapping potential], the 1D system is unconditionally stable even
without the trapping potential, though in a latter case it forms a
droplet (see, e.g., \cite{inf_1d_bose}).

Ueda and Leggett \cite{ul} have risen a question of whether the
system of 1D attractive bosons in a finite-size rotating toroidal
trap can remain irrotational, that is demonstrate the behavior
typical for a superfluid -- the so-called  Hess-Fairbank (HF)
effect \cite{hessfairbank}. On the basis of their (approximate)
treatment, they arrived at the conclusion that the HF effect
should take place in the system, provided the strength of
interaction is below a certain threshold value. However, recently
Berman {\it et al}. \cite{smerzi} have questioned this result,
arguing that the HF effect disappears at arbitrarily small value
of attractive interaction due to a specific quantum instability.

In this Section, we resolve the above-mentioned controversy by an
exact numeric study of $N$ rotating 1D bosons with the
$\delta$-functional attractive interaction. We do observe the HF
effect predicted by Ueda and Leggett (though the threshold
interaction differs by a factor of 2 from the value found in
Ref.~\cite{ul}). Moreover, our data clearly demonstrate that in
the large-$N$ limit the conventional methods of weakly interacting
Bose gas -- like Gross-Pitaevskii (GP) equation\cite{GP} and
Bogoliubov technique \cite{nnb} -- become applicable, the
dimensionless small parameter controlling the accuracy of the
approximation being just $1/N$. The classical-field language of GP
equation renders the issue of the presence and disappearance of
the HF effect especially transparent. The effect persists as long
as the condensate density remains uniform, and disappears with
breaking the spatial homogeneity of the condensate. In a few
particle system, the deviations from the mean-field picture are
significant. In particular, with decreasing $N$ the HF effect
gradually becomes indistinguishable from the generic effects of
angular momentum quantization.

\begin{figure}
\begin{center}
\includegraphics[width=8cm]{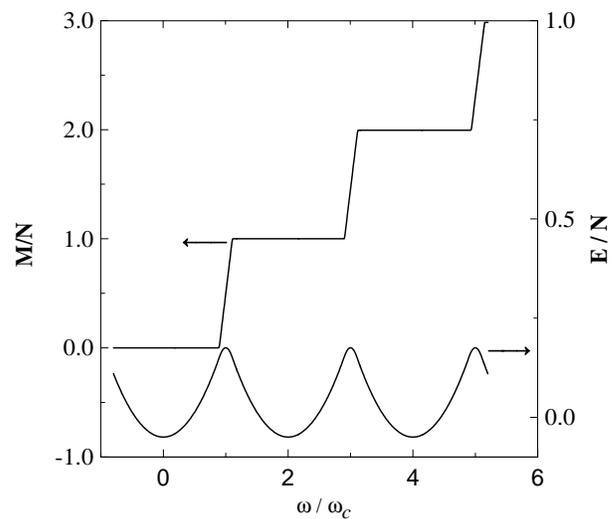}
\end{center}
\caption{Dependence of the groundstate energy $E$ (lower curve)
and angular momentum $M$ (upper curve) on the rotation frequency
$\omega$. $\gamma=0.1$, $N \to \infty$.  } \label{staircase}
\end{figure}

Consider $N$ bosons of the mass $m$ placed in the rotating torus
of radius $R$ and cross-sectional area $S = \pi r^2$. At low
enough temperature and with the condition $r \ll R$ met, the
system is quasi-one-dimensional, and the effective 1D Hamiltonian
in the rotating (with angular frequency $\omega$) frame reads
\begin{equation}
 H = \sum _k
 {\left( k - \frac{\omega}{2} \right) ^2
n^{\mathstrut}_k}
 + \frac{g}{2} \sum _{k,l,q}
 { a^{+}_{k}  a^{+}_{l}
  a^{\mathstrut}_{l - q}  a^{\mathstrut}_{k + q}} \; ,
  \label{H}
\end{equation}
where the integers $k,l,q$ stand for angular momenta, $a^{+}_{k}$
creates a boson with angular momentum $k$, $n^{\mathstrut}_k =
a^{\mathstrut +}_k  a^{\mathstrut}_k $; $g = 2 a / m R S $ (with
$a<0$ the 3D scattering length) is the effective vertex of pair
interaction; we use units $\hbar=1$ and $\omega_c = 1 $, where
$\omega_c  = 1 / 2 m R^2$ is the critical rotation frequency equal
to the period of variation of the groundstate energy as a function
of $\omega$. The groundstate of the system is defined by the three
parameters, $N$, $\gamma = |g|(N-1)$, and $\omega$.

Ueda and Leggett \cite{ul} have analyzed the model (\ref{H}) in
the Hartree-Fock approximation in the Fock basis of angular
momentum eigen states $\{ \left| ... , n_{-1}, n_0, n_1, ...
\right> \}$. They argued that in the limit of $\gamma \to 0$, when
no more then two single-particle angular momentum  eigen modes
survive, their approximation is superior with respect to the other
treatments of weakly interacting bosons. A typical result for
$\gamma \ll 1$ is shown in Fig.~\ref{staircase}. 
The HF effect -- plateaus in the
angular momentum curve, $M(\omega)$, -- takes place almost at any
$\omega$, except for a close vicinity of the critical frequency
$\omega_c$. With increasing $\gamma$, the size of the plateaus
gets smaller, and the HF effect completely disappears at some
critical point $\gamma=\gamma_c \sim 1 $. At this point we note
that  $\gamma_c$ cannot be found accurately with the treatment of
Ref.~\cite{ul}, because more than two single-particle angular
momentum eigen modes are involved in the formation of the
groundstate. This is immediately seen from the {\it variational}
Hartree-Fock treatment, when all the particles are placed into one
and the same spatially dependent single-particle state $\psi_0(x)$
(non-uniform Bose-Einstein condensate), the wavefuction
$\psi_0(x)$ being defined from the minimal energy condition, which
leads to the GP equation
\begin{equation}
\left(i \partial / \partial x + \omega /2 \right)^2 \psi_0
 - 2\pi \gamma |\psi_0|^2 \psi_0 - \mu \psi_0 = 0 \; ,
\label{GPE}
\end{equation}
where $\mu$ is taken to satisfy the normalization condition $\int
| \psi_0(x)|^2 \, {\rm d} x = 1$. Solving Eq.~(\ref{GPE}) reveals
the nature of the HF effect, as well as the mechanism of its
disappearance. At $\omega \leq \omega_*(\gamma)$, with
$\omega_*(\gamma)$ satisfying
\begin{equation}
 \left(\omega_* - 2k\right) ^2 = 1 - 2 \gamma \; ,
\label{square}
\end{equation}
the solution $\psi_0(x)$ is uniform and thus irrotational. At
$\omega > \omega_*(\gamma)$, the rotational symmetry of the
problem breaks down: The density becomes non-uniform, and  the
rotation of the density profile gives rise to the increase of the
angular momentum with growing $\omega$. At $\gamma
> \gamma_c=1/2$ the minimal-energy solution is non-uniform even
without the rotation \cite{ueda_dn}, and the HF effect totally
disappears.

A simple and instructive way of arriving at the relation
(\ref{square}) is the Bogoliubov treatment \cite{nnb} for the
elementary excitation spectrum in assumption of the uniform
condensate. Considering the energy, $\epsilon_1$, of the first
excited state, one observes that at $\omega > \omega_*$, the
uniform condensate is {\it thermodynamically} unstable:
$\epsilon_1 < 0$. At $\gamma>1/2$  the energy $\epsilon_1$ becomes
imaginary, indicating {\it dynamical} instability of the
homogeneous solution at any $\omega$.

At $\gamma \gg 1$ the solution of Eq.~(\ref{GPE}) is strongly
non-uniform and corresponds to a rotating condensate
droplet -- bright soliton \cite{inf_1d_1964,hashimoto}.

\begin{figure}
\begin{center}
\includegraphics[width=8cm]{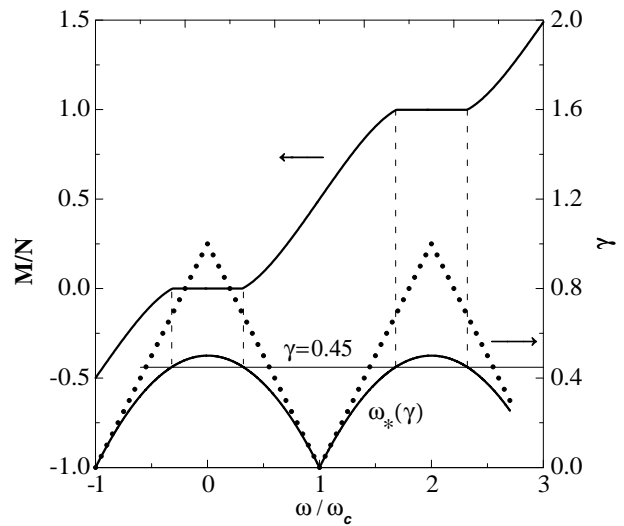}
\end{center}
\caption{
 Function $M(\omega)$ at $\gamma=0.45$,
 as predicted by GP equation (upper solid curve). Lower solid curve represents
$\omega_*(\gamma)$, according to Eq.~(\ref{square}).  Dotted curve
is the prediction of Ref.~\cite{ul} for $\omega_{*} (\gamma)$.}
\label{twographs}
\end{figure}

\begin{figure}
\includegraphics[width=8cm]{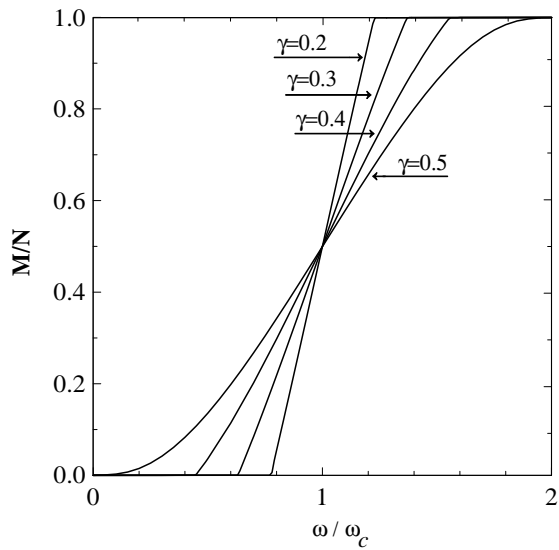}
\caption{Gross-Pitaevskii equation results for the function
$M(\omega)$ at different $\gamma$'s.} \label{lines}
\end{figure}

To obtain the GP prediction for the angular momentum as a function
of $\omega$, we solved Eq.~(\ref{GPE}) by numerically minimizing
GP energy functional, see Figs.~\ref{twographs},~\ref{lines}. At
$\gamma \ll 1$, the GP results coincide, up to higher-order
corrections, with those of Ref.~\cite{ul}. At $\gamma \sim 1$,
however, deviations become significant.

\subsection{Simulation for finite $N$}

As we have already mentioned, the very existence of the HF effect
(and thus the applicability of the weakly interacting gas
treatments) has been questioned recently by Berman {\it et al}.
\cite{smerzi}. This conclusion of Ref.~\cite{smerzi} seems rather
strange and counter-intuitive in view of the known exact results
for the 1D attractive bosons in the non-restricted geometry. The
small parameter that guarantees applicability of the mean-field
approach is just the inverse number of particles
\cite{inf_1d_bose}. {\it A priori} we do not see how the finite
system size can qualitatively change the situation.

In view of this controversy, as well as keeping in mind the fact
that mesoscopic results are interesting in their own value,  we
performed an exact numeric study of the model (\ref{H}) for
different numbers of bosons.

Analytic summation of diagonal kinks contribution
due to 
point interaction $U_{pqrs} = U_0 \delta_{p+q,r+s}$
increases full energy by constant
$\Delta E = U_0 (2 N^2 - N)$, and particle
energy obtains term depending on occupation
$\tilde \epsilon_k = \left( k - \omega/2 \right)^2
- U_0 n_k$.

After typical update, the weight of MC configuration
acquires the factor $\sim e^{-\delta E (\tau-\tau_0)}$.
Main difference from lattice models
is that $\delta E$ can be unlimitedly large,
so with large $\delta E$, the weight of new configuration
would be too small to appear in simulation
and the efficiency of the update approaches zero.
Therefore we must choose for update different places
with different probability
to make simulation effective enough.

Note, for old $\epsilon_k = \left( k - \omega/2 \right)^2$,
there exist simple relation $\delta E = 2 q(q+\Delta)$
with $\Delta$ characterizing the place of update.
Therefore we can neglect second term $-U_0 n_k$ 
in new particle energy $\tilde \epsilon$,
and
 choose $q$ using simple analytics,
given in Appendix \ref{details_1d} for all types of updates.
Though for small $k$, both terms in new particle energy
$\tilde \epsilon_k$ are comparable 
(for in this model $U_0 = \gamma/2(N-1)$, $\gamma \le 1$),
this trick is aimed 
mostly for resonable choosing of large momentum $k$
where $\langle n_k \rangle \simeq 0$ 
making second term rather virtual.

\subsection{Results}

\begin{figure}
\begin{center}
\includegraphics[width=8cm]{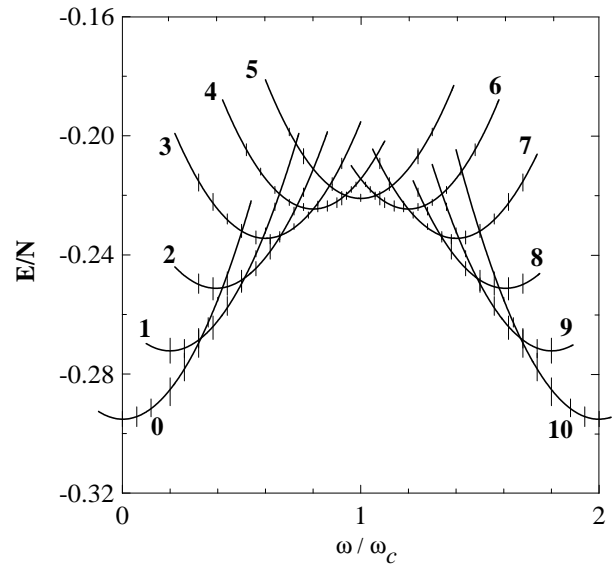}
\end{center}
\caption{ Groundstate energies in different angular momentum
sectors for $N=10$ at $\gamma = 1/2$. The bars are the Monte Carlo
data with errors. The solid curves are the parabolic fits.
Integers stand for the values of the angular momenta.} \label{e10}
\end{figure}

\begin{figure}
\begin{center}
\includegraphics[width=8cm]{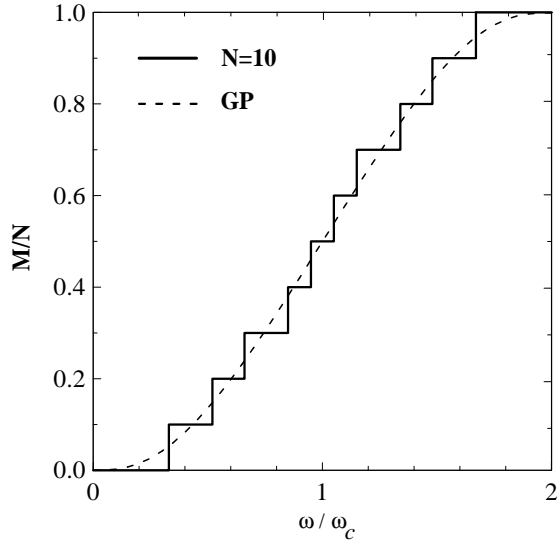}
\end{center}
\caption{Groundstate angular momentum $M$ as a function of the
rotation frequency $\omega$ for $N=10$ at $\gamma = 1/2$ (solid
line). Absolute error is on the order of $10^{-2}$. Dotted curve
is the $N \to \infty$ limit obtained by solving Gross-Pitaevskii
equation. } \label{rotation10}
\end{figure}

We traced the evolution of the system properties with $N$ variying
from 2 to 100, at different $\omega$'s and $\gamma$'s. The
simplest characteristics that we studied was the groundstate
energy, which we calculated by simulating the groundstates in
different angular momentum sectors, with subsequent selecting the
global minimum, see Fig.~\ref{e10}. This procedure yields also the
curve $M(\omega)$. For a finite-$N$ system this curve is
essentially stepwise due to the quantization of angular momentum.
Comparing this curve at large enough $N$ to the Gross-Pitaevskii
solution, we find an excellent agreement, see
Fig.~\ref{rotation10}.

To quantitatively trace the difference between GP and Monte Carlo
results, we compared corresponding answers for the groundstate
energies at different $N$'s. We found that starting from $N
\approx 10$ the deviation between the two results scales as $1/N$,
which confirms that in the $N \to \infty$ limit the
Gross-Pitaevskii equation yields a perfect description of the
groundstate properties.

\begin{figure}
\begin{center}
\includegraphics[width=8cm]{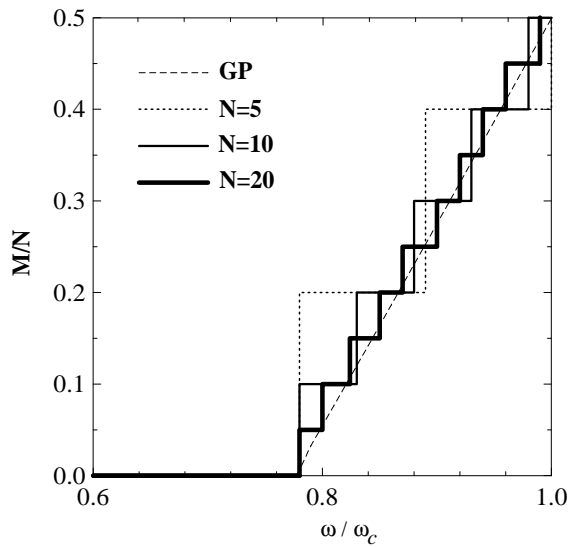}
\end{center}
\caption{ Groundstate angular momentum $M$ as a function of
rotation frequency $\omega$  for $N=5,10, 20$ at $\gamma = 0.2 <
\gamma_c=1/2$. Absolute error is on the order of $10^{-2}$. The
dashed line is the result of the Gross-Pitaevskii equation.
  } \label{m_5_10_20}
\end{figure}

In Fig.~\ref{m_5_10_20} we present the $M(\omega)$ curves for
$\gamma=0.2 < \gamma_c=1/2$, at different particle numbers. Once
again we see an excellent agreement with the GP equation at large
$N$. The HF plateau is unambiguously revealed. Note, that at $N
\lesssim 5$ the HF plateau at zero momentum is indistinguishable
(by its size) from the rest of the quantized-momentum plateaus.

\begin{figure}
\begin{center}
\includegraphics[width=8cm]{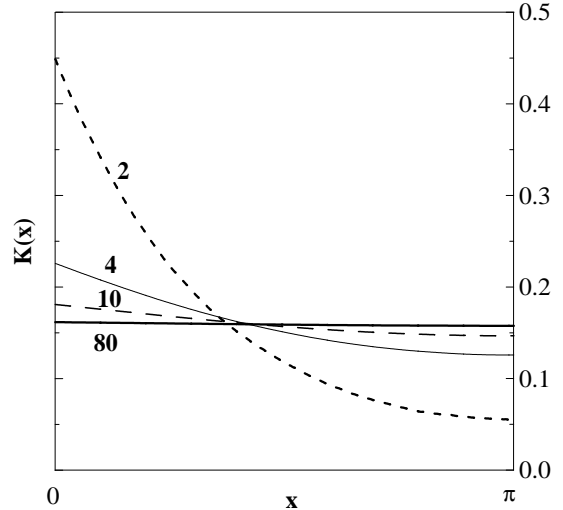}
\end{center}
\caption{Density-density correlator $K(x)$ ($\omega = 0$) at
$\gamma = 0.25 < \gamma_c=1/2$ for $N=80,10,4,2$. The correlations
grow up with decreasing $N$.} \label{korr_below}
\end{figure}

\begin{figure}
\begin{center}
\includegraphics[width=8cm]{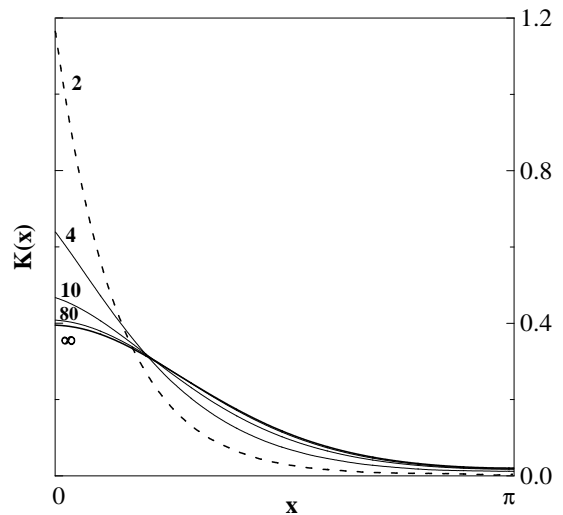}
\end{center}
\caption{Density-density correlator  $K(x)$ ($\omega = 0$) at
$\gamma = 0.75 > \gamma_c=1/2$  for $N=80,10,4,2$. Solid curve
corresponds to the  Gross-Pitaevskii equation. The particle
positions are correlated even in the macroscopic limit indicating
the non-uniformity of the groundstate.} \label{korr_above}
\end{figure}

To get an insight into the inner structure of the ground state, we
calculate the two-particle densityn correlator $K(x)=
\left<\Psi^+(x')\Psi^+(x'+x)\Psi(x'+x)\Psi(x')\right>_{x'}/N(N-1)$.
In Figs. \ref{korr_below} and \ref{korr_above} we present $K(x)$ (
$\omega = 0$) for $\gamma_1 = 0.25 $ and $0.75 $, that is for
uniform and non-uniform (in the macroscopic limits) cases,
respectively. At small enough $N$ there is no qualitative
difference between the two cases. At large $N$ the difference is
clearly seen. Once again note an excellent agreement with the GP
equation.

Summarizing, we developed a novel Quantum Monte Carlo
algorithm based on the generic principles
of the Diagrammatic Monte Carlo approach \cite{ctwl}.
The algorithm 
samples exact diagrammatic expansion (in terms of the pair
interaction) of the imaginary-time evolution operator in the 
{\it momentum} representation.
In comparison with usual real-space methods,
this algorithm is more efficient 
in simulation of weakly interacting systems.
Moreover, it can be applied to
the models which can not
be studied by real-space methods.
The sign problem is weaker here
making possible to reach temperatures low enough
to study ground state.
 In the case of attractive pair potential, all diagrams are
positive-definite and the method is very efficient, allowing to
simulate up to 100 and more particles.

Using exact Quantum Monte Carlo algorithm,
the groundstate properties of one-dimensional bosons
with attractive $\delta$-functional interaction in a rotating
toroidal trap are studied.
The Hess-Fairbank effect -- absence of a response to the
trap rotation -- is observed in a certain area of the parameter
space. The fact that in the $N \to \infty$ limit the
Gross-Pitaevskii equation yields a perfect description of the
groundstate properties is proved.

The author is grateful to Profs. V.A. Kashurnikov, B.V. Svistunov,
and N.V. Prokof'ev for drawing my attention to this problem and
numerous helpful discussions. This work was supported by the
Russian Foundation for Basic Research.

\appendix

\section{Weighting the time of new kink}
\label{Z1}
With adding new kink in time $\tau$
in range $( \tau_{min} \dots \tau_{max} )$,
the statistical weight of MC configuration
receives a factor $ \sim e^{-Q(\tau - \tau_{min})}$,
where $Q$ denotes change of worldline energy
in the range $(\tau_{min} \dots \tau)$
after update.

Time $\tau$ is determined with probability
$\Delta \tau \frac{e^{-Q(\tau - \tau_{min})}} {Z_1(Q, \tau_{max} - \tau_{min})}$
using relation
\begin{equation}
\nonumber
 \tau = \tau_{min} + \mathrm{P}(Q, \tau_{max} - \tau_{min}, R), 
\end{equation}
where $R$ is the random number
uniformly distributed between 0 and 1.

For convenience, we define the following functions:
\begin{equation}
\label{z1_def}
Z_1(Q,T) =
 \begin{cases}
	\frac{e^{-QT} - 1}{ -Q }, &\text{ usually}, \\
	T, & \text{ if } |Q| \to 0, \\
	\frac{1}{Q}, & \text{ if } Q \to +\infty, \\
	\frac{e^{-QT}}{-Q}, & \text{ if } Q \to -\infty,
 \end{cases}
\end{equation}
\begin{equation}
\label{OnePoint_def}
\mathrm{P}(Q,T,R) =
 \begin{cases}
	-\frac{ \ln{(1 + R(e^{-QT} - 1))} }{Q}, &\text{ usually}, \\
	RT, & \text{ if } |Q| \to 0, \\
	\frac{1}{Q}\ln{R}, &\text{ if } Q \to +\infty,\\
	T - \frac{1}{|Q|}\ln{R}, &\text{ if } Q \to -\infty.
 \end{cases}
\end{equation}
Limiting cases should be realized separately
to avoid algorithm inefficiency
and precision loss.

\section{Updates in details}
\label{details_1d}

\subsubsection{Creation/Annihilation of Two Kinks}

\begin{figure}
\begin{center}
\includegraphics[width=6cm]{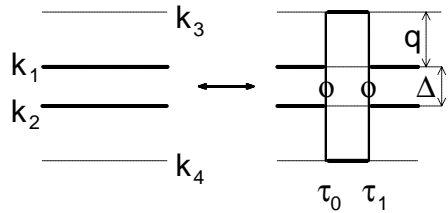}
\end{center}
\caption{Creation/Annihilaton of Two Kinks.}
\label{figFAF}
\end{figure}

This process is shown in Fig.~\ref{figFAF}.

Two kinks are created as follows:
\begin{enumerate}
\item{}
Time of first kink $\tau_0$ from $0$ to $\beta$ is chosen
in random way with probability $\frac{\Delta \tau}{\beta}$.
\item{}
Two not-empty world lines are found going through $\tau_0$
with momenta $k_1$ and $k_2$,
including possible case $k_1 = k_2$. Let us denote
the probability to choose these world lines  $W(k_1, k_2)$.
\item{}
The second kink will be created at the time $\tau_1$
determined using Appendix \ref{Z1}.
\begin{equation}
\nonumber
	\tau_1 = \tau_0 + \mathrm{P}(2q(q+\Delta), \tau_{max} - \tau_0).
\end{equation}
With increasing $k_3$ and $k_4$, 
their average occupation becomes exponentially small,
so $\tau_{max}$ remains intact.
This fact allows us to take into account 
all possible $k_3, k_4$ in single process.
The analytics for choosing $q$ is given
in Appendix \ref{Analytics},

\item{}
The probability to accept this process is found
from the balance equation (\ref{FAF_balance}).

\end{enumerate}

The scheme of annihilation:

\begin{enumerate}
\item{}
A random kink
having parameters $\tau_0$, $k_1$, $k_2$, $q$, $U_{eff}$, $n_{eff}$
is taken with probability
$\frac{1}{Q_{new}} = \frac{1}{Q_{old}+2}$.
\item{}
The kink chosen and its right neighbour
(the direction is fixed to remove polyvalence)
must form the ``kink-antikink'' pair.
\item{}
The probability to choose $q$
in the direct process is $\frac{W(q)}{Z_3}$.
\item{}
The annihilation probability $p _\leftarrow ^{(acc)}$
of this pair is determined from the balance equation
(\ref{FAF_balance}).
\end{enumerate}

\begin{eqnarray}
\label{FAF_balance}
W_{old} W(k_1,k_2) \frac{W(q)}{Z_3} \frac{\Delta \tau} {\beta} \frac{\Delta \tau e^{(\dots)}}{Z_1} p_\rightarrow^{(acc)} = \\
  W_{old} (-\Delta \tau U_{eff}  n_{eff} )^2 e^{(\dots)}  \frac{1}{Q+1} p_\leftarrow^{(acc)} \nonumber.
\end{eqnarray}

\begin{figure}
\begin{center}
\includegraphics[width=6cm]{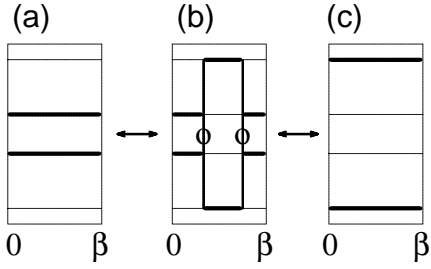}
\end{center}
\caption{About possibility to annihilate a pair of kinks
in two ways resulting in different configurations.}
\label{FAF_closed}
\end{figure}

(Pay attention to the possibility of pair annihilation
in two different ways when all four momenta $k_1, ..., k_4$
are untouched by other kinks, see Fig.~\ref{FAF_closed}.
To remove this duality we disable the annihilaton
when $\tau_1 < \tau_0$).

\subsubsection{Kink Expanding/Contracting - particle version}

\begin{figure}
\begin{center}
\includegraphics[width=6cm]{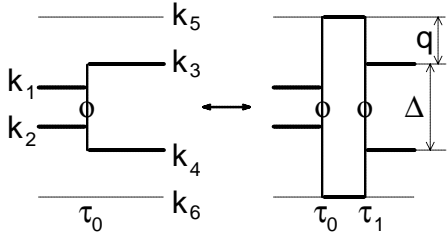}
\end{center}
\caption{Process of Kink Expanding/Contracting - particle version.}
\label{fig122}
\end{figure}

The direct process (see Fig.~\ref{fig122}) is done as follows:

\begin{enumerate}
\item
A random kink having parameters $\tau_0$, $k_1, ..., k_4$,
$U_1$, $n_1$ is chosen with probability $\frac{1}{Q}$.
\item
Choice of $q$ and $\tau_1$ is done similarly to the creation
of two kinks with probabilities $W(q)/Z_3$ and
$\exp{(- \Delta E (\tau_1 - \tau_0))} / Z_1$, respectively.
\item
The probability to accept this update
is determined using balance equation
(\ref{balance_122}).
\end{enumerate}

Inverse transformation:

\begin{enumerate}

\item
A random kink having parameters $\tau_0$, $k_1, ..., k_4$,
$U_2$, $U_3$, $n_2$, $n_3$ is chosen with probability
$\frac{1}{Q_{new}} = \frac{1}{Q_{old}+1}$.
\item
The kink chosen must, with its right neighbour, form the pair 
which can be contracted into single kink.
\item
The probability to choose this $q$ in direct process
equals $W(q)/Z_3$.
\item
The probability to to accept this update
is determined from the balance equation (\ref{balance_122}).
\end{enumerate}

\begin{eqnarray}
\label{balance_122}
W_{common}  (-\Delta \tau U_1 n_1) \frac{1}{Q}  \frac{W(q)}{Z_3} \frac{\Delta \tau e^{(\dots)}}{Z_1}   p_\rightarrow^{(acc)} = \\
  W_{common} (-\Delta \tau)^2 U_2 U_3 n_2 n_3 e^{(\dots)} \frac{1}{(Q+1)} p_\leftarrow^{(acc)} \nonumber
\end{eqnarray}

\subsubsection{Kink Expanding/Contracting - ``hole'' version}

In contrast to previous update,
this process needs momenta $k_5$, $k_6$
to be occupied.
However, in each moment $\tau_0$
only limited set of occupied pairs $(k_5, k_6)$ exists
with $k_5+k_6 = k_1+k_2$ 
(see Fig.~\ref{fig122hole}),
i.e. fitting this update.
Therefore the momenta should not be weighted
using formulas of Appendix \ref{Analytics}.
Otherwise, we choose a random pair
from rather small set.

\begin{figure}
\begin{center}
\includegraphics[width=6cm]{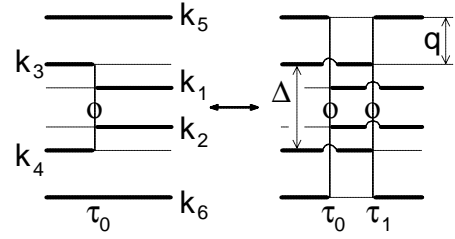}
\end{center}
\caption{Kink Expanding/Contracting - ``hole'' version.}
\label{fig122hole}
\end{figure}

The direct update is made as follows:
\begin{enumerate}
\item
A random kink having parameters
$\tau_0$, $k_1, ..., k_4$, $U_1$, $n_1$
is chosen with probability $\frac{1}{Q}$.
\item
All pairs $(k_5, k_6)$ fitting this process are determined;
for each pair, the parameters 
$\tau_{max}$, $n_2$, $n_3$, $U_2$, $U_3$, $\Delta E$
are found.
\item
 Each pair $(k_5, k_6)$ is chosen with probability
$W(k_5, k_6) \sim U_2 U_3 n_2 n_3 Z_1( \Delta E, \tau_{max} - \tau_0)$;
\item
The balance equation (\ref{balance_hole122})
is used to determine the probability to accept this update.
\end{enumerate}

The inverse process is made as follows:
\begin{enumerate}
\item
A random kink having parameters $\tau_0$,
$k_1$, $k_2$, $k_5$, $k_6$, $U_2$, $U_3$, $n_2$, $n_3$
is chosen with probability
$\frac{1}{Q_{new}} = \frac{1}{Q_{old}+1}$.
\item
The kink chosen and its right neighbour must form the pair 
which can be contracted into single kink
fitting the direct ``hole'' update.
\item
The probability $W(k_5, k_6)$ to choose this pair
$(k_5, k_6)$ in direct process, is determined.
\item
The probability to accept this update is determined 
from the balance equation (\ref{balance_hole122}).
\end{enumerate}

\begin{eqnarray}
\label{balance_hole122}
W_{common}  (-\Delta \tau U_1 n_1) \frac{1}{Q}  W(k_5,k_6) \frac{e^{(\dots)}}{Z_1}   p_\rightarrow^{(acc)} = \\
  W_{common} (-\Delta \tau)^2 U_2 U_3 n_2 n_3 e^{(\dots)} \frac{1}{(Q+1)} p_\leftarrow^{(acc)} \nonumber
\end{eqnarray}

\subsubsection{Entangling of a Kink with a Worldline}

\begin{figure}
\begin{center}
\includegraphics[width=6cm]{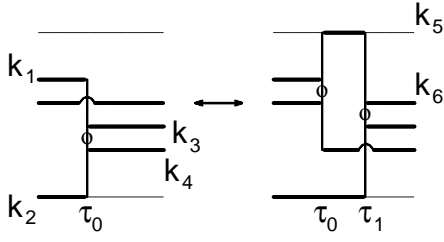}
\end{center}
\caption{
Entangling of a Kink with a Worldline.
In the case shown here, the upper parts of left and right
kinks are touched.
}
\label{fig_o122}
\end{figure}

This process can not be made of the processes
described above.
More visually, this is the only process able to change
the winding numbers of worldlines. 
There exist many versions of this update.
The case touching upper parts of left and right kinks
is shown in Fig.~\ref{fig_o122},

The momentum $k_6$ should be occupied at $\tau_0$,
thus there is only limited set of pairs $(k_5, k_6)$
fitting the process of entangling.
Therefore, the schemes of direct and inverse updates
are similar to that for the previous transformation.
The only correction should be made
is to choose upper or lower parts of kinks,
if present.

\subsubsection{Shifting the Kink Through Another}

\begin{figure}
\begin{center}
\includegraphics[width=6cm]{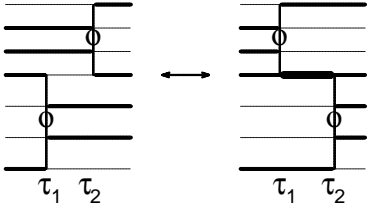}
\end{center}
\caption{Shifting the Kink Through Another.
Kinks exchange their times;
the occupation numbers are changed for worldlines
with momenta common for both kinks.}
\label{figShTh}
\end{figure}

Carrying out shift in time, one must take into account 
possible collisions with neighbouring kinks.
While simple shift is enough for simulating
Fermi systems, Bose case should incorporate
shift "through" neughbouring kink (Fig.~\ref{figShTh}).
In this process, the kinks are exchanged by time:
$\tau_2^{new} = \tau_1^{old}$,
$\tau_1^{new} = \tau_2^{old}$.

\begin{figure}
\begin{center}
\includegraphics[width=6cm]{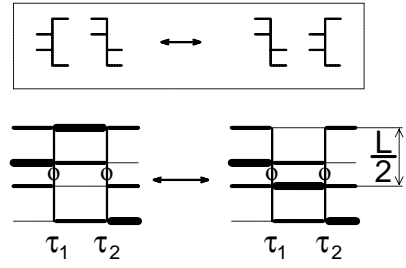}
\end{center}
\caption{
To the process of Shifting the Kink Through Another.
In simulation of periodical chain,
two kinks can occupy all the same momenta
though not forming the ``kink-antikink'' pair.
$L$ denotes the length of a chain.
}
\label{figShTh_4moms}
\end{figure}

The number $K$ of common momenta can be 1 to 4.
The case $K=4$ is usually associated
with the ``kink-antikink'' pair, so the 
shifting through the kink
can be replaced by the combination of
Annihilation and Creation.
However, another case 
shown in Fig.~(\ref{figShTh_4moms})
can appear in simulation of Bose chain.

Due to symmetry, direct and inverse processes are identical.
The scheme is as follows:

\begin{enumerate}
\item
A random kink having parameters $\tau_0$, $k_1, ..., k_4$
is chosen with probability $\frac{1}{Q}$;
\item
The nearest kink on the right with at least one the same
momentum, is found;
\item
The parameters $\Delta E_{before}$, $\Delta E_{after}$,
 $n_{before}$, $n_{after}$ are determined;
\item
The probability to accept the update
is determined from the balance equation
(\ref{balance_ShTh}).
\end{enumerate}

\begin{eqnarray}
\label{balance_ShTh}
W_{common} U_1 U_2 n_{before} e^{-\Delta E_{before} (\tau_2-\tau_1)} \frac{1}{Q} p_\rightarrow^{(acc)} = \\
  W_{common} U_1 U_2 n_{after} e^{-\Delta E_{after} (\tau_2-\tau_1)} \frac{1}{Q} p_\leftarrow^{(acc)}, \nonumber
\end{eqnarray}
or, finally,
$$
\alpha \equiv \left| \frac{p_\rightarrow^{(acc)}}{p_\leftarrow^{(acc)}} \right| \\
 = \frac{n_{after}}{n_{before}} e^{-(\Delta E_{after} - \Delta E_{before})(\tau_2-\tau_1)}.
$$

\section{Choosing momentum analytically}
\label{Analytics}

\begin{figure}
\begin{center}
\includegraphics[width=8cm]{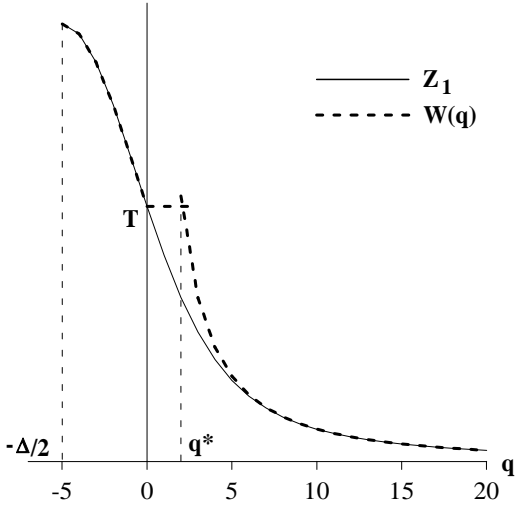}
\end{center}
\caption{$Z_1(T, 2q(q+\Delta))$ as a function of $q$ (solid curve)
and its piecewise approximation W(q) (dotted curve)
used to choose $q$ analytically.}
\label{z1fig}
\end{figure}

We use the fact $\delta E = 2 q ( q + \Delta ),$
where $\Delta$ is fixed in choosing the place of action,
and $ q > - \Delta / 2$.
The value of $Z_1 (\delta E(q), T)$ от
as a function of $q$, is shown in Fig.~\ref{z1fig}.
It can be approximated by following piecewise function
\begin{equation}
\label{zq}
Z_1 \approx W(q) =
\begin{cases}
 Z_1 (E(q), T),  &\text{if } q < 0 \\
 const = T,  &\text{if } 0 <q \le q^ \star,  \\
\frac{1}{ E(q)},  &\text{if } q > q^ \star,
\end{cases}
\end{equation}
making possible to determine $q$ analytically.

The choice of $q^ \star $ is based 
on the relation  $1/E(q^ \star) \approx T$:

\begin{equation}
q^* = 
\begin{cases}
 \sqrt{\frac{1}{2T}},  &\text{if }\Delta = 0, \\
 \frac{\Delta}{2} \left( \sqrt{1 + \frac{2}{T \Delta ^2}} - 1 \right),  &\text{if } \Delta > 0 .
\end{cases}
\end{equation}

The possibility to choose given $q$ equals $\frac{W(q)}{Z_3}$
where $Z_3 = Z^{(-)} + Z^{(0)} + Z^{(+)}$,
and
\begin{eqnarray}
 Z^{(-)} = \sum\limits_{q=q_{min}}^{-1} {W(q)}, \\
 Z^{(0)} = \sum\limits_{q > 0}^{q^ \star} {T} = T q^ \star, \\
 Z^{(+)} \simeq \sum\limits_{q>q^\star}^{\infty} {\frac{1}{2q(q + \Delta)}} \simeq \nonumber \\ 
 \begin{cases}
	\frac{1}{2q^{\star}} &\text{if } \Delta=0, \\
	\frac{1}{\Delta} ln\left(1 + \frac{\Delta}{q^{\star}}\right) &\text{if } \Delta>0
 \end{cases}
\end{eqnarray}
correspond to three pieces of approximation (\ref{zq}).
Values $Z^{(-)}$, $Z^{(0)}$, $Z^{(+)}$
define boundary values $ R^{(-)} = \frac{Z^{(-)}}{Z_3}$
and $R^{(+)} = 1 - \frac{Z^{(+)}}{Z_3}$.

To choose $q$, the random number $R$
uniformly distributed between 0 and 1.
 is taken.
Initially, we determine the range
($q<0$, $0 < q \le q^\star$, $q > q^\star $),
then $q$ according to the range found.
When $R \le R^{(-)}$, we find the value $q<0$,
making $\sum \limits_{q'=q_{min}}^{q'<q} {W(q')}$
greater than $R Z_3$.
Other cases allow analytical relation:
$q = 1 + q^\star \frac{R - R^{(-)}}{R^{(+)} - R^{(-)}}$
if $R^{(-)} < R \le R^{(+)}$, and
\begin{equation}
\nonumber
q =
 \begin{cases}
  \frac{q^\star}{r}, &\text{ при } \Delta = 0, \\
  \frac{\Delta }{ \left( 1 + \Delta/{q\star} \right)^r - 1}, &\text{ при } \Delta > 0, \\
 \end{cases}
\end{equation}
where $r = \frac{R - R^{(+)}}{1 - R^{(+)}}$,
if $R > R^{(+)}$.

The scheme modifications for cases
$\Delta \ge -1$ (without range $q<0$)
and $q^\star = 0$ (without range $0 < q \le q^{\star}$)
are straightforward.

{\bf Note}. 
In summation $Z^{(-)}$, the care should be taken
in managing overflow.
The exponential index should not be limited by any value.
Otherwise the relations between values of $Z_1$
would be broken, making most probable momenta $q<0$
with nonequal energies $(q - \omega/2 \omega_c)^2$
to have similar weights.

Some way to avoid this, is to decrease the energy difference
 $\delta E = E^{new} - E^{old} = 2q(q + \Delta)$
by some value into safe region.

\end{document}